# AI and Legal Argumentation:
# Aligning the Autonomous Levels of AI Legal Reasoning

**Dr. Lance B. Eliot**
Chief AI Scientist, Techbruim; Fellow, CodeX: Stanford Center for Legal Informatics
Stanford, California, USA

**Abstract**

Legal argumentation is a vital cornerstone of justice, underpinning an adversarial form of law, and extensive research has attempted to augment or undertake legal argumentation via the use of computer-based automation including Artificial Intelligence (AI). AI advances in Natural Language Processing (NLP) and Machine Learning (ML) have especially furthered the capabilities of leveraging AI for aiding legal professionals, doing so in ways that are modeled here as *CARE*, namely Crafting, Assessing, Refining, and Engaging in legal argumentation. In addition to AI-enabled legal argumentation serving to augment human-based lawyering, an aspirational goal of this multi-disciplinary field consists of ultimately achieving autonomously effected human-equivalent legal argumentation. As such, an innovative meta-approach is proposed to apply the Levels of Autonomy (LoA) of AI Legal Reasoning (AILR) to the maturation of AI and Legal Argumentation (AILA), proffering a new means of gauging progress in this ever-evolving and rigorously sought domain.

## 1 Background on Legal Argumentation

Legal scholars and the legal profession overall have expended much effort and attention toward the formulation and analysis of legal argumentation [5] [31] [32] [42]. This focus on legal argumentation has entailed in-depth research and conceptual examination of what constitutes legal argumentation all told, along with the detailing of the day-to-day practical aspects concerning how to best undertake legal argumentation and how to gauge when legal argumentation is being accomplished well or performed poorly [10] [41] [46].

Taking a brief side tangent for purposes of definitional clarity, there is pervasive ambiguity about the meaning of the phrase "legal argumentation" as to its exact definition and denotations. Likewise, the phrase "legal argument" also has varied connotations. To make this discussion relatively parsimonious, this paper considers that "legal argumentation" shall herein refer to the broadest scope of all facets involved in the act of argumentation within the field of law. Meanwhile, the phrase "legal argument" will be reserved for use when discussing a particular instance or sub-element within the umbrella of legal argumentation. It is hoped that this will enable the discussion to be more readily cohesive and consistent, doing so without any loss of substantive indication or impact. Even if the proposed demarcation is not of satisfaction to some, or could be seen as arguable unto itself, it nonetheless does not undercut or distill the essence of the discussion. Given that brief clarification, the discussion can now continue with that caveat so noted.

Legal argumentation is reasonably viewed as a variant of generic argumentation [18] [68], suggesting that there is a macroscopic realm entailing "argumentation" of any kind or nature and that legal argumentation is an instantiation or incarnation of argumentation into the domain of law. Thus, there is the full spectrum of generic argumentation, out of which there could be argumentation applied in specific disciplines, such as the law, medicine, engineering, and the like.

Some controversy exists over the question of whether generic argumentation is any different fundamentally from the argumentation utilized within any particular domain or discipline. In other words, generic argumentation might be simply reused within the field of law, for example, and not be structurally or intrinsically any different due to the advent of its use in the legal domain.

There are ongoing debates as to legal argumentation being essentially a subset of generic argumentation or whether it can be construed as a superset. If one asserts that legal argumentation goes beyond the realm of generic argumentation, presumably then it is the case that legal argumentation is a superset. If one asserts that legal argumentation is no more than the application of generic argumentation into the nuances of the law, presumably legal argumentation is a subset thereof.

Whichever side of the debate one takes, it does appear that the efforts to explore, extend, and mature our understanding of generic argumentation provides insights for legal argumentation, and similarly that the efforts underlying the maturation of legal argumentation can be funneled into generic argumentation comprehension. As such, this paper takes no side on this question and merely notes its importance, synergistic effect, and ongoing consideration in such matters.

One could sensibly claim that legal argumentation is more than merely one amongst equals in terms of the instantiation of generic argumentation. Of all the various disciplines that draw upon argumentation, legal argumentation has a most notable history and need for pushing ahead on extending what argumentation consists of and how it can be described and utilized.

Legal argumentation is a vital cornerstone of justice, underpinning an adversarial form of law. This is not to imply that legal argumentation does not appear in and nor serves a crucial role in other forms of law, and merely is noted as especially required and to some degree revered when placed into an adversarial architecture. The adversarial approach places legal argumentation at the front and center of legal matters and therefore heightens interest and priority to the state of legal argumentation.

To bolster legal argumentation, there have been efforts to dovetail the use of computer-based systems into the field of and the everyday acts of legal argumentation. In its simpler form, computers can be a storehouse from which legal argumentation can be undertaken by humans, and otherwise, the computer is serving as a modest aid in a rather mechanistic manner.

The field of Artificial Intelligence (AI) has long sought to intermix with the field of Legal Argumentation, doing so to try and extend the computer capabilities to be a more powerful aid to humans involved in legal argumentation. Indeed, there is extensive research that has attempted to augment or undertake legal argumentation via the use of computer-based automation including especially the AI subfields of Natural Language Processing (NLP), Machine Learning (ML), and Knowledge-Based Systems (KBMS).

An overarching aspiration for AI in the legal argumentation realm would be to have the computer or machine be able to autonomously perform some or all of the aspects of legal argumentation. Ultimately, achieving autonomously effected human-equivalent legal argumentation is a long-sought goal for those in this specialty of AI. Within the legal field, there has been a similar interest, and thus there is a decidedly multi-disciplinary effort involved in these matters. For those interested in technological capabilities, the attention tends to go toward developing AI that can perform legal argumentation. Also, there is a rightful concern and attention toward the societal implications of AI that could perform legal argumentation, including whether this would be considered a form of practice-of-law, and what implications this has for those seeking legal advice and those proffering legal advice.

Returning to the synergistic aspects earlier mentioned, there is an ongoing dance, as it were, entailing the synergistic effects of generic argumentation with legal argumentation, and the synergistic effects of AI with both generic argumentation and legal argumentation. To emphasize, it is not as though AI is somehow a sideline aspect that is solely attempting to implement argumentation (of any kind). Instead, AI also contributes to the understanding of what argumentation consists of, and thus probes and provides substantive advancement to argumentation, whether it be for generic argumentation or the field of legal argumentation.

There is, in a sense, plenty of synergies to go around. AI as a field is inherently itself enriched by seeking to be an element of and contributor to generic argumentation and also of legal argumentation. This occurs on both a theoretical level of deriving new theories and conceptual gains and also on a practical basis of deploying usable systems. In short, it would seem that each field of inquiry is apt to gain by the deployment of the other into its focal of inquiry.

This paper seeks to contribute to that aspiration and does so by proposing an innovative meta-approach toward applying the Levels of Autonomy (LoA) of AI Legal Reasoning (AILR) to the maturation of AI and Legal Argumentation (AILA), proffering a new means of gauging progress in this ever-evolving and rigorously sought domain.

Doing so provides a well-needed means of measuring progress in the aspiration of integrating AI and legal argumentation. Researchers and scholars can make use of the LoA AILR to AILA to assess progress in the theoretical and conceptual efforts of integration, while practitioners can assess the strengths, weaknesses, and progress in the deployment of such capabilities. This paper also provides a model known as *CARE*, namely Crafting, Assessing, Refining, and Engaging in legal argumentation, which outlines the core facets by which AI is integrated into legal argumentation.

In Section 1 of this paper, the topic of legal argumentation is further introduced and addressed. Doing so establishes the groundwork for the subsequent sections. Section 2 introduces the Levels of Autonomy (LoA) of AI Legal Reasoning (AILR), which is instrumental in the discussions undertaken in Section 3. Section 3 provides an indication of the AI-enablement of legal argumentation as it applies to the LoA AILR. The final section, Section 4, covers additional considerations and recommendations.

This paper then consists of these four sections:
- Section 1: Background on Legal
    Legal Argumentation
- Section 2: Autonomous Levels of
    AI Legal Reasoning
- Section 3: Legal Argumentation and
    AI Enablement
- Section 4: Additional Considerations and
    Future Research

## 1.1 Research on Legal Argumentation and AI

As mentioned, legal argumentation is crucial to the field of law, and this is emphasized in the case of resolving legal disputes [9]: "Argumentation is particularly important in law: a legal case typically centres on a conflict between two parties which is resolved by each side producing arguments in an effort to persuade the judge that their side is right. The judge then decides which party to favour, and publishes a decision in which he argues why his decision is justified. Modelling legal reasoning can then be seen, to a large extent, in terms of modelling argument, and so it is unsurprising that attempts to understand legal argumentation have been a key strand in AI and Law."

Of course, legal argumentation has numerous other roles besides solely dealing with particular legal conflicts or disputes and permeates a wide array of legal activities and efforts. In addition, legal argumentation pervades the adversarial process [3]: "Legal reasoning usually takes place in the context of a dispute between adversaries, within a prescribed legal procedure. This makes the setting inherently dynamic and multiparty and raises issues of strategy and choice. For example, there is work on optimal strategies for adversaries in debates with an adjudicator, given their preferences over the possible outcomes of a debate and their estimates of what the adjudicator will likely accept."

There is much discourse regarding whether there is a potential unification theory that might someday be postulated to fully provide a complete semblance of legal argumentation in the law, of which there is doubt cast that such an all-encompassing theory is viable, as will be indicated next.

Consider the pluralistic model of law, which can be depicted as [40]: "I accept the term 'pluralistic' for this descriptive model of legal argument because it reflects the fact that law arises from value choices made by different persons at different times and it acknowledges that there are different ways to determine what choices they made. Heads of administrative agencies, judges, legislators, and the people all make law. There are a variety of methods for interpreting the law they have made, and as a result our interpretations of the law are sometimes contradictory." And as further stated: In their view, the law is not essentially a unitary system that can be explained by a 'grand unifying theory' or 'foundational analysis.' Instead, our system of law is characterized by the fact that multiple legitimate forms of legal arguments exist."

Some are seeking to mathematically or computationally model legal argumentation in the quest of seemingly embodying the law in an axiomatic way. There are serious questions expressed about the efficacy of these approaches in terms of making a key assumption that law is ultimately determinate and calculable. For example, as stated by [40]: "For pluralists, law is inherently indeterminate because valid but contradictory legal arguments potentially exist regarding the interpretation of the law."

Efforts to utilize rule-based non-monotonic logic-based systems for modeling legal argumentation are also called into question [56]: "We also argued that 'traditional' rule-based nonmonotonic logics (whether argumentation-based or not) are of limited use, and that the role of cases, principle, purpose and value should not be ignored, as well as the importance of dynamics, procedure and multi-agent interaction. This holds for the law but also for related areas such as policy making, group decision making and democratic deliberation. More generally, legal applications of logic confirm the recent trend of widening the scope of logic from deduction to information flow, argumentation and interaction."

Part of this can be attributed to what some suggest is a false belief that there are legal specifications that if only found or voiced would then allow for AI to be codified accordingly, as mentioned in [33]: "Although it is commonly accepted that legal decisions must be justified in a rational way, there are hardly explicit legal specifications as to what the justification should consist of. One of the important problems in the study of legal argumentation is which standards of legal soundness the argumentation should meet."

According to [33], it is vital to realize that legal argumentation consists of several components, involving philosophical, theoretical, reconstruction, empirical, and practice facets: "The philosophical component attends to the normative foundation of a theory of legal argumentation. In the theoretical component, models for legal argumentation are developed, in which the structure of legal argument and norms and rules for argument-acceptability are formulated. The reconstruction component shows how to reconstruct legal argument in an analytical model. The empirical component investigates the construction and evaluation of arguments in actual legal practice. Finally, the practical component considers how various results forwarded by the philosophical, theoretical, analytical, and the empirical components might be used in legal practice."

This though does not summarily discount the value of mathematical formulation for legal argumentation, noting that one should not necessarily discard the advantages out-of-hand and that such computational models can be of significant benefit [9]: "It is, of course, the case that similar issues underpin one well-established and highly-developed theory: that of formal logic and mathematical proof. It is no coincidence that much of the formal computational treatment of argumentation has its roots in ideas developed from AI inspired contributions to logic and deductive reasoning. So one finds in mathematical proof theory core concepts such as: precisely defined means for expressing assertions (e.g. formulae in a given logical language); accepted bases on which to build theorems (e.g. collections of axioms); procedures prescribing the means by which further theorems may be derived from existing theorems and axioms (e.g. templates for inference rules); and precise concepts of termination (e.g. a sentential form is derivable as a theorem, 'true'; or is logically invalid, 'false')."

But legal argumentation still has extraordinary facets that require going beyond traditional mathematical formulations, of which the intrinsic element that legal argumentation necessitates of *persuasion* opens the matter further, accordingly [9]: "One can summarise the distinction between argumentation and proof by the observation that the object of argumentation is to persuade (to acceptance of a given claim; to performance of a desired action, and so on). Unlike the concept of 'proof'—at the level of deriving a sentential representation of an assertion—whether an argument is "correct" is not a factor, and, indeed, "correctness" may not even be sensibly defined. In contrast, mathematical reasoning, in order to have any value, must be correct where 'correctness' has a strict, formal definition: beyond this requirement, however, notions of "persuasiveness" are unimportant."

Perhaps this is succinctly stated by the assertion in [49]: "But there is more to a legal argument than reasoning or logic. Justices and attorneys bring an interpretive context, argumentative and rhetorical strategies, and other more general models of the domain and the world to an oral argument."

A distinction made by some researchers is that legal argumentation is not a pure logic-based effort and should be seen as one of more so rhetoric-based [40]: "Legal reasoning is not composed of deductive arguments framed for the purpose of proving the truth of a particular proposition but is a species of rhetoric designed to persuade others to accept a particular interpretation of the law. But how is the persuasiveness of a legal argument to be evaluated? What is the yardstick against which we measure the 'correctness' of legal reasoning? There are two fundamental types of challenges to legal arguments: 'intramodal' and 'intermodal' challenges. Intramodal critiques challenge legal arguments on their own terms, while intermodal critiques address the validity or weight to be accorded to each type of argument."

Per the research of [39], persuasiveness is indeed vital to legal argumentation: "Legal reasoning entails a practice of argumentation. The reasons given for the conclusions reached are to be measured by their persuasiveness, not by reference to some established true state of affairs."

Part of the conundrum underlying this ingredient of persuasiveness is that besides being difficult if not indeterminable to model, there are canons of construction related to the law that can essentially land on either side of a legal argumentation case. For example, Llewellyn made this point about the nature of opposing canons [45]: "As in argument over points of case-law, the accepted convention still, unhappily requires discussion as if only one single correct meaning could exist. Hence there are two opposing canons on almost every point. An arranged selection is appended. Every lawyer must be familiar with them all: they are still needed tools of argument."

Some examples of these opposing canons can be illustrative of this matter. Consider this one [45]:

"1. A statute cannot go beyond its text."

And this corresponding and opposite one:

"1. To effect its purpose a statute text may be implemented beyond its text."

Similar examples in [45] demonstrate this same facility of the law and legal argumentation:

"2. Statutes in derogation of the common law will not be extended by construction."

"2. Such acts will be liberally construed if their nature is remedial."

And this one about common law [45]:

"3. Statutes are to be read in the light of the common law and a statue affirming a common law rule is to be construed in accordance with the common law."

"3. The common law gives way to a statue which is inconsistent with it and when a statue is designed as a revision of a whole body of law applicable to a given subject it supersedes the common law."

And so on, including even the role of punctuation [45]:

"24. Punctuation will govern when a statue is open to two constructions."

"24. Punctuation marks will not control the plain and evident meaning of the language."

For adversarial legal reasoning, it has been postulated that there are three levels involved, namely the logic level, argument level, and dialogical level [55]: "Adversarial legal reasoning consists of three levels, the logic level (generates arguments), argument level (organizes argument, identifies attack relations, ascertain acceptability of the arguments for given points in a debate), the dialogical level (how arguments can be deployed in a dispute)." And that this can form the basis for constructing a legal argumentation framework [55]: "A three-level model is proposed, where a formal argumentation framework is built around a logical system and itself embedded in a dialectical protocol for dispute, in such a way that, each time a party adds or retracts information, the argumentation framework reassesses the resulting state of the dispute."

A notable aspect brings up the role of attacks related to legal argumentation. It is insufficient to undertake legal argumentation without also considering the importance and essence of attacks too. Attempts to model the variants of attack include formulations such as denial of the premises, alternative action for same effect, side effects of an action, and so on. In work by [36], the details of fifteen forms of attack are described, though this is a starter list and the researchers mention as such: "However, we consider only the opening stages of an argument, leaving counter-attacks, shifts in the burden of proof, and similar issues to later work."

Research by Dung [18] has highlighted the importance of attacks in legal argumentation: "Roughly, the idea of argumentational reasoning is that a statement is believable if it can be argued successfully against attacking arguments. In other words, whether or not a rational agent believes in a statement depends on whether or not the argument supporting this statement can be successfully defended against the counterarguments. Thus, the beliefs of a rational agent are characterized by the relations between the 'internal' arguments supporting his beliefs and the "external" arguments supporting contradictory beliefs."

In a means of thinking about legal argumentation, the adversarial approach has perhaps fueled or further bolstered the complexity of ordinary argumentation, as outlined by [56]: "All these aspects of the law, i.e., its orientation to future and not fully anticipated situations, the tension between the general terms of the law and the particulars of a case, and the adversarial nature of legal procedures, make that legal reasoning goes beyond the literal meaning of the legal rules and involves appeals to precedent, principle, policy and purpose, and involves the attack as well as the construction of arguments. A central notion then in the law is that of argumentation."

There is also the need to realize that multiple agents or actors are potentially involved in legal argumentation. This in turn rachets up the complexities [56]: "From this analysis it appears that different kinds of agents, with different functions engage in legal reasoning in different contexts: the addressees of the norms (the citizens), the producers of the norms (the legislators), the appliers of the norms (the judges and administrators), and the enforcers of the norms (the administrators/police officers). The reasoning forms employed in the law may thus not only depend on the nature of the issue addressed but also on the context in which the reasoning takes place."

AI researchers tend to describe argumentation models as being classified or known as COMMA (Computational Models of Argument), and that a cornerstone includes the use of Argumentation Mining (AM) [37]: "The goal of argumentation mining, an evolving research field in computational linguistics, is to design methods capable of analyzing people's argumentation." Furthermore, AM is still an evolving field [37]: "Despite the lack of an exact definition, researchers within this field usually focus on analyzing discourse on the pragmatics level and applying a certain argumentation theory to model and analyze textual data at hand." There is also the use of Argument Markup Language (AML), such as in the case of the Araucaria system, leveraging elements of XML accordingly: "The argument markup language (AML) defines a set of tags that indicate delimitation of argument components (loosely, propositions), tags that indicate support relationships between those components, and tags that indicate the extent of instances of argumentation schemes. The design of AML builds on results in the theory of argument and its application in AI, and therefore, although similar in spirit, is significantly different from earlier attempts."

Per earlier emphasis, there is not as yet any unifying or unitary theory that has been explicated that could shore-up these varied methods and approaches [69]: "As yet, there is no unitary theory of argumentation that encompasses the logical, dialectical, and rhetorical dimensions of argumentation and is universally accepted. The current state of the art in argumentation theory is characterized by the coexistence of a variety of theoretical perspectives and approaches, which differ considerably from each other in conceptualization, scope, and theoretical refinement."

No singular discipline or specialty is likely to gain notable ground if it does not remain open to other allied disciplines in the cross-boundary multi-headed tentacles of legal argumentation [33]: "The study of legal argumentation draws its data, assumptions and methods from disciplines such as legal theory, legal philosophy, logic, argumentation theory, rhetoric, linguistics, literary theory, philosophy, sociology, and artificial intelligence. Researchers with different backgrounds and from various traditions are attempting to explain structural features of legal decision-making and justification from different points of view."

Recall too that legal argumentation is not an island unto itself and that it presumably in one way or another inextricably entangled with generic argumentation. Note this point about persuasion in argumentation as mentioned in 1917 by Ketcham [44]: "The art of persuading others to think or act in a definite way. It includes all writing and speaking which is persuasive in form." And, going back to 1898 and the work of MacEwan [47]: "Argumentation is the process of proving or disproving a proposition. Its purpose is to induce a new belief, to establish truth or combat error in the mind of another."

These underlying elements are acknowledged in today's AI effort towards encompassing even generic argumentation embodiment, since any AI, even that outside of the legal realm and for any nature of human intelligence purposes, must facilitate argumentation [70]: "The field of artificial argumentation plays an important role in Artificial Intelligence research. The reason for this is based on the recognition that if we are to develop robust intelligent machines able to act in mixed human-machine teams, then it is imperative that they can handle incomplete and inconsistent information in a way that somehow emulates the way humans tackle such a complex task."

As stated concisely [70]: "Humans argue. Machines should be able to argue too if we aim to achieve mixed teams in a hybrid society."

Turns out that the legal field, because of its intrinsic requirement for argumentation, provides a valuable testbed for developing AI overall, as revealed in a review of the AI field in this regard [56]: "As can be seen from this review, the development of logical models of legal reasoning nowadays proceeds mostly within AI & law and is very much driven by real examples and applications. We think that the latter is a fortunate development, since it shows that the law is a rich testbed for AI theories of reasoning and argument."

Upon furtherance of this point, the richness of legal argumentation and its keystone role in law are added benefits to the attempts at applying AI [56]: "Law is of vital importance to society, promoting justice and stability and affecting many people in important aspects of their private and public life. Creating and applying law involves information processing, reasoning, decision making and communication, so the law is a natural application field for artificial intelligence. While AI could be applied to the law in many ways (for example, natural-language processing to extract meaningful information from documents, data mining and machine learning to extract trends and patterns from large bodies of precedents), the fact that law is part of society makes logic particularly relevant to the law."

And, continuing [56]: "Since law has social objectives and social effects, it must be understood by those affected by it, and its application must be explained and justified. Hence the importance of clarity of meaning and soundness of reasoning, and hence the importance of logic for the law and for legal applications of AI. This review aims to introduce AI researchers to the law as a rich testbed and important application field for logic-based AI research, with a particularly marked concern for logical models of legal argument."

One subtle but significant aspect of legal argumentation is that it cannot normally be done in a hidden or obscured manner, namely that the expectation is that legal argumentation will be explainable and interpretable. As mentioned in [7]: "Legal decisions must be justified. In law, the answer is not enough: the reasons for the answer must be given in order to guide future decisions, to ensure consistency of decisions, and to attempt to persuade the losing side of why they lost, perhaps leading to acceptance of the decision."

Another consideration is the defeasible nature of the law and thus its impact upon legal argumentation [7]: "Law is defeasible. Legal rules can be overturned by finding that an exception applies, or by finding a conflicting law, or by distinguishing the case so that the rule does not apply."

Recent efforts of AI applied to argumentation include the IBM Project Debater system [5], which makes use of argument clustering, argument summarization, and the mapping of arguments to key points underlying an argumentation. In addition, when coping with a new situation involving an unrehearsed argumentation session, there is the use of "first principles" of reusing prior debates or prior arguments to enact a de novo argument synthesis [10].

Do not though somehow be led into imagining that argumentation is nearly solved, which it is most decidedly not, and likewise nor is legal argumentation.

As aptly stated in [9]: "Most significantly, however, that the body of theory, techniques, and applications we have discussed is very far from encompassing a final, definitive description of the scope and limits of what argumentation-based approaches can offer to the furtherance of AI as a scientific discipline: many questions remain unresolved, many avenues unexplored, and many applications offer a wealth of possibilities for future work. When people participate in reasoned debate they are engaging in argumentation not demonstration. Thus argumentation, rather than logical demonstration, should be seen as the core technique for justifying claims."

There is a dearth of AI-enabled legal argumentation systems for today's use by the legal profession [3]: "While the theoretical advances on models of legal argument have been impressive and a number of valuable prototype systems have been developed, no systems have been deployed in everyday practice yet."

A somewhat daunting concern is that the ability to have AI fully operate in any substantive autonomous manner as to legal argumentation might be equivalent to the AI problem in its entirety of achieving human intelligence capacities [3]: "Developing artificial tools that capture the human ability to argue is an ambitious research goal, and it may ultimately prove to be as difficult as developing AI in general."

Meanwhile, various attempts at structuring legal argumentation and the modeling of legal argumentation continue, of which some are noted next.

One viewpoint is that there are three pillars involved in AI argumentation facets: "The three pillars of the development of argumentation-enhanced intelligent machines are, from my point of view: (i) modeling and reasoning on socio-cognitive components like trust using computational models of argument which are able to deal with incomplete and conflicting information, (ii) mining argument structures in natural language text to detect, e.g., potential fallacies, recurrent patterns, and inner strength, and (iii) analyzing and understating the role of emotions in real-world argumentative situations (e.g., debates) to inject such information in the computational models of argument to better cast incomplete and inconsistent information when emotions play a role."

One indication is that there are three approaches overall [33] "In the past 30 years three more or less consistent approaches to legal argumentation can be distinguished: the logical, the rhetorical and the dialogical approach."

As recently pointed out in [3], the earlier work by Dung was especially insightful since it illuminates the importance of abstraction in legal argumentation: "Dung's insight was that arguments could be abstract and this freed them from any particular method of generation, whether using a particular logic, particular argument schemes, or case based methods. Once in the framework all arguments were equal. This separated consideration of the status of arguments from the logic that produced them."

In exploring how humans undertake legal argumentation, the work by Huhn [40] postulates that there are five types of legal argumentation: "Five types of legal argument exist: text, intent, precedent, tradition, and policy. Each type of argument may be considered an information set or a category of evidence admissible to prove what the law is." And these are respectively based on differing concepts of the law and justice [40]: "Each type of legal argument is based upon a different conception of justice; that is, a different source of the law. The first four types of legal argument are of ancient lineage, while the fifth, policy analysis, has been expressly acknowledged as a valid legal argument only in the twentieth century."

In terms of what is meant by the fifth type, the policy analysis, here's the indication provided [40]: Policy analysis proceeds in two steps: a predictive statement and an evaluative judgment. The court first predicts the consequences that will flow from giving the law one interpretation or another and then decides which set of consequences is more consistent with the underlying values of the law. In attacking a legal argument based on policy analysis, one may challenge either the predictive statement of consequences or the evaluative judgment. Policy analysis can be contrasted with each of the foregoing sources of law. Rather than requiring the court to ascertain the value choices made by others, policy analysis invites the court itself to make a policy choice by balancing all of the relevant values and interests affected by the decision to pursue a particular policy."

Finally, before concluding this section of the paper, it is perhaps noteworthy to consider an ongoing question underlying law schools and the training of lawyers, which offers insights into the nature of legal reasoning and human intelligence, all of which are essential to the discussion of legal argumentation.

Per the provocative commentary in [40]:

> "Students enter law school expecting to learn 'the law,' that is, rules of law. They conceive law to be a science, a set of determinate rules that govern human behavior. Moreover, students are frustrated when law professors insist the principal purpose of legal education is not learning rules of law, but rather learning 'to think like lawyers.' What exactly does it mean 'to think like a lawyer?' 'To think like a lawyer' is to be adept at legal analysis; it is to be able to predict, argue, and decide what the law is in hard cases. The purpose of legal education is to train students in the mastery of this skill."

The next section of this paper introduces the autonomous levels of AI Legal Reasoning, doing so to then aid Section 3 then explores how legal argumentation automation varies across the levels of autonomy. Section 3 also covers more introspection of the AI intertwining with legal argumentation. Section 4 provides some conclusionary remarks and also an indication of recommended future research.

## 2 Autonomous Levels of AI Legal Reasoning

In this section, a framework for the autonomous levels of AI Legal Reasoning is summarized and is based on the research described in detail in Eliot [24].

These autonomous levels will be portrayed in a grid that aligns with key elements of autonomy and as matched to AI Legal Reasoning. Providing this context will be useful to the later sections of this paper and will be utilized accordingly.

The autonomous levels of AI Legal Reasoning are as follows:

Level 0: No Automation for AI Legal Reasoning
Level 1: Simple Assistance Automation for AI Legal Reasoning
Level 2: Advanced Assistance Automation for AI Legal Reasoning
Level 3: Semi-Autonomous Automation for AI Legal Reasoning
Level 4: Domain Autonomous for AI Legal Reasoning
Level 5: Fully Autonomous for AI Legal Reasoning
Level 6: Superhuman Autonomous for AI Legal Reasoning

### 2.1 Details of the LoA AILR

See **Figure A-1** for an overview chart showcasing the autonomous levels of AI Legal Reasoning as via columns denoting each of the respective levels.

See **Figure A-2** for an overview chart similar to Figure A-1 which alternatively is indicative of the autonomous levels of AI Legal Reasoning via the rows as depicting the respective levels (this is simply a reformatting of Figure A-1, doing so to aid in illuminating this variant perspective, but does not introduce any new facets or alterations from the contents as already shown in Figure A-1).

### 2.1.1 Level 0: No Automation for AI Legal Reasoning

Level 0 is considered the no automation level. Legal reasoning is carried out via manual methods and principally occurs via paper-based methods.

This level is allowed some leeway in that the use of say a simple handheld calculator or perhaps the use of a fax machine could be allowed or included within this Level 0, though strictly speaking it could be said that any form whatsoever of automation is to be excluded from this level.

### 2.1.2 Level 1: Simple Assistance Automation for AI Legal Reasoning

Level 1 consists of simple assistance automation for AI legal reasoning.

Examples of this category encompassing simple automation would include the use of everyday computer-based word processing, the use of everyday computer-based spreadsheets, the access to online legal documents that are stored and retrieved electronically, and so on.

By-and-large, today's use of computers for legal activities is predominantly within Level 1. It is assumed and expected that over time, the pervasiveness of automation will continue to deepen and widen, and eventually lead to legal activities being supported and within Level 2, rather than Level 1.

### 2.1.3 Level 2: Advanced Assistance Automation for AI Legal Reasoning

Level 2 consists of advanced assistance automation for AI legal reasoning.

Examples of this notion encompassing advanced automation would include the use of query-style Natural Language Processing (NLP), Machine Learning (ML) for case predictions, and so on.

Gradually, over time, it is expected that computer-based systems for legal activities will increasingly make use of advanced automation. Law industry technology that was once at a Level 1 will likely be refined, upgraded, or expanded to include advanced capabilities, and thus be reclassified into Level 2.

### 2.1.4 Level 3: Semi-Autonomous Automation for AI Legal Reasoning

Level 3 consists of semi-autonomous automation for AI legal reasoning.

Examples of this notion encompassing semi-autonomous automation would include the use of Knowledge-Based Systems (KBS) for legal reasoning, the use of Machine Learning and Deep Learning (ML/DL) for legal reasoning, and so on.

Today, such automation tends to exist in research efforts or prototypes and pilot systems, along with some commercial legal technology that has been infusing these capabilities too.

### 2.1.5 Level 4: Domain Autonomous for AI Legal Reasoning

Level 4 consists of domain autonomous computer-based systems for AI legal reasoning.

This level reuses the conceptual notion of Operational Design Domains (ODDs) as utilized in the autonomous vehicles and self-driving cars levels of autonomy, though in this use case it is being applied to the legal domain [19] [20] [21].

Essentially, this entails any AI legal reasoning capacities that can operate autonomously, entirely so, but that is only able to do so in some limited or constrained legal domain.

### 2.1.6 Level 5: Fully Autonomous for AI Legal Reasoning

Level 5 consists of fully autonomous computer-based systems for AI legal reasoning.

In a sense, Level 5 is the superset of Level 4 in terms of encompassing all possible domains as per however so defined ultimately for Level 4. The only constraint, as it were, consists of the facet that the Level 4 and Level 5 are concerning human intelligence and the capacities thereof. This is an important emphasis due to attempting to distinguish Level 5 from Level 6 (as will be discussed in the next subsection)

It is conceivable that someday there might be a fully autonomous AI legal reasoning capability, one that encompasses all of the law in all foreseeable ways, though this is quite a tall order and remains quite aspirational without a clear cut path of how this might one day be achieved. Nonetheless, it seems to be within the extended realm of possibilities, which is worthwhile to mention in relative terms to Level 6.

### 2.1.7 Level 6: Superhuman Autonomous for AI Legal Reasoning

Level 6 consists of superhuman autonomous computer-based systems for AI legal reasoning.

In a sense, Level 6 is the entirety of Level 5 and adds something beyond that in a manner that is currently ill-defined and perhaps (some would argue) as yet unknowable. The notion is that AI might ultimately exceed human intelligence, rising to become superhuman, and if so, we do not yet have any viable indication of what that superhuman intelligence consists of and nor what kind of thinking it would somehow be able to undertake.

Whether a Level 6 is ever attainable is reliant upon whether superhuman AI is ever attainable, and thus, at this time, this stands as a placeholder for that which might never occur. In any case, having such a placeholder provides a semblance of completeness, doing so without necessarily legitimatizing that superhuman AI is going to be achieved or not. No such claim or dispute is undertaken within this framework.

## 3 Legal Argumentation and AI Enablement

In this section, the advent of AI and legal argumentation is explored in several respects. First, a model coined as CARE for Create, Assess, Refine, and Engage is introduced and provided as context for AI Legal Argumentation (AILA) facets. The Toulmin formalism for argumentation is next showcased and discussed, including augmentation that considers the importance of attack vectors, horizontal layers, vertical layers, and recursion. This is followed by an examination of hard cases versus clear cases, providing a four-square quadrant for case difficulty comparison purposes. A persuasion cloud that represents a legal argumentation search space is next shown and used to illustrate the process of identifying a winning argument all told. Finally, typing together the Section 1 and Section 2 discussions, a chart is provided that indicates the maturation of AI Legal Argumentation along with the Levels of Autonomy (LoA) of AI Legal Reasoning (AILR).

A series of figures are included in the discussions to aid in illustrating the matters addressed.

### 3.1 AI Legal Argumentation and CARE

A model coined as CARE for Create, Assess, Refine, and Engage is shown in **Figure B-1**. This reflects the key activities or processing that an AI-enabled Legal Argumentation (AILA) system would be expected to perform. The wording indicates:

- Create a Legal Argument
- Assess a Legal Argument
- Refine a Legal Argument
- Engage in Legal Argumentation

As earlier indicated in Section 1, the phrasing involving the word "argument" is construed at both a narrow perspective such as an element within a larger overall argument, and can also be interpreted as the totality of the argument that is being forged.

In brief, an AILA is expected to be able to craft anew a legal argument, though this should be understood as not necessarily implying "from scratch" per se. In other words, a new argument might readily be based on prior arguments and not originated from thin air, as it were. Thus, the notion of a new argument is one that is being created anew versus for example examining an existing argument for purposes of assessing the argument. In fact, an additional portion of CARE involves the assessment of a legal argument. This might be undertaken during the crafting of a new argument, or it might occur when seeking to discover counterarguments as part of an adversarial methodology, and so on. Furthermore, it is anticipated that an AILA would be able to refine a legal argument, receiving as input an existing legal argument, and attempt to refine it to bolster its strengths and reduce its weaknesses. Finally, an AILA is expected to be able to engage in a dialogue regarding a legal argumentation.

Note that each of these elements of CARE is anticipated to be interleaving with each other and though listed as seemingly distinct activities or processes are to be considered as intermixing and interoperative.

The CARE model can be aligned orthogonally to the various stages or layers typified in artificial argumentation, such as the research in [3]: "We consider the following five main layers: structural, relational, dialogical, assessment, and rhetorical. Structural layer: How are arguments constructed? Relational layer: What are the relationships between arguments? Dialogical layer: How can argumentation be undertaken in dialogues? Assessment layer: How can a constellation of interacting arguments be evaluated and conclusions drawn? Rhetorical layer: How can argumentation be tailored for an audience so that it is persuasive?"

### 3.2 Toulmin's Formalism of Argumentation

One of the most commonly cited formalisms for argumentation consists of the argument structure identified by Toulmin [68]. This argument structure is frequently utilized as a core foundation for establishing theories and practical tools for legal argumentation.

As an example of the reuse of Toulmin's argument structure, consider this research by Marshall [49]: "As our representational starting point, we used Toulmin's formalism for logical structure. According to his scheme, a datum is some fact or observation about the situation under discussion that leads to some further observation or fact, the claim. The relation between the two is characterized by a rule of inference, a warrant, that serves to link the information set forth in the datum and claim. A backing supports the warrant with some knowledge structure from the argument's domain. A Toulmin argument structure may also include various kinds of qualifications of the claim (qualifiers) and allow for exceptions (rebuttals). The categories provided by this structure are useful for expressing portions of argument logic."

Here is a rationale for making use of Toulmin's argument structure [49]: "What does this work suggest about the essential elements of a tool to support the formulation, organization, and presentation of arguments? First, it suggests that we need a system of representations that captures reasoning and allows it to be structured by interpretive information. Toulmin structures and our current system of representations to organize reasoning are a good start. Toulmin structures can also function as the input to reasoning analysis mechanisms such as assumption-based truth maintenance systems. Second, we need a solid understanding of the formulation process."

**Figure B-2** indicates the core elements of Toulmin's argument structure.

**Figure B-3** augments the core elements of Toulmin's argument structure by emphasizing the importance of attack vectors associated with each element, along with an overarching attack collective. As earlier indicated, robust legal argumentation entails the need to consider attacks that can be waged.

**Figure B-4** showcases that the Toulmin argument structure should be considered as multi-dimensional. A legal argument of any substance is likely to have a multitude of horizontal layers and also have vertical layers. These layers contain sub-arguments. Since there is a dependency of the sub-arguments to the encompassing argument(s), it is crucial that the sub-arguments also be given due consideration. There is a recursive facility involved in deriving sub-arguments, such that there are sub-arguments within sub-arguments, and so on.

### 3.3 Four-Square Grid Case Difficulty Comparison

In **Figure B-5**, a four-square grid of quadrants representing case difficulty comparison is presented.

Along the vertical axis are the two major classifications of legal cases, consisting of hard cases and so-called clear cases. A clear case is denoted as one that is relatively routine, readily commonly understood, and previously experienced (some refer to these as "soft" cases). Hard cases are considered arduous to legally extricate and resolve, are usually one-off's that have not previously been experienced, etc. [38].

Along the horizontal axis is the use of a conventional argument versus an unconventional argument. When combined with the two major classifications of hard cases and clear cases, a four-square grid or set of quadrants is established and can be utilized accordingly. In particular:

- Hard Case – Conventional Argument: *Unconvincing (Deficient)*
- Hard Case – Unconventional Argument: *Inventive (Formative)*
- Clear Case – Conventional Argument: *Convincing (Expected)*
- Clear Case – Unconventional Argument: *Distended (Exorbitant)*

This set of indications is insightful for AILA as to the differences between forming legal argumentation in the instance of a hard case versus a clear case. Of course, it is not necessarily apparent as to whether a given case is a hard case or a clear case until sufficient assessment has been undertaken.

### 3.4 Legal Argumentation Cloud Search Space

In Figure B-6, a diagram is used to represent the search space involved in examining legal arguments.

This is a cloud in the sense that it is a potentially large search space that needs to be established and utilized by an AILA. While seeking to identify a "winning argument" (based on persuasiveness and other attributes), there is some $n$ number of possible legal arguments that can be potentially (exhaustively) identified. There is a need by the AILA to winnow the search space to gauge those legal arguments $m$ that are considered in the winning realm. These are contextually dependent and time-dependent.

## 3.5 Legal Argumentation and LoA AILR

As shown in **Figure B-7**, it is useful to align the evolution of AI-enablement of Legal Argumentation (AILA) with the Levels of Autonomy (LoA) of AI Legal Reasoning (AILR).

For each of the levels of autonomy of AI Legal Reasoning, the impacts upon legal argumentation will be distinctive. A keyword phrasing is used in Figure B-7 to indicate these impacts and consists of:

**LoA AILR – Legal Argumentation**

Level 0: *n/a*

Level 1: *Mechanistic (Low)*

Level 2: *Mechanistic (High)*

Level 3: *Expressive*

Level 4: *Domain Fluency*

Level 5: *Full Fluency*

Level 6: *Meta-Fluency*

In brief, at Level 0, which consists of no automation for AI Legal Reasoning, the applicability to legal argumentation is considered not applicable ("n/a"), simply due to the by-definition that there is no AI involved at this level. At Level 1, simple assistance automation, the characterization is indicated as "Mechanistic (Low)" since the AI is abundantly unrefined and limited in any substantive practical capacity to the legal argumentation advent, thus considered mechanistic. At Level 2, advanced assistance automation, the characterization is indicated as "Mechanistic (High)" since the AI at this level can modestly assist in legal argumentation but is considered quite preliminary in doing so.

At Level 3, the first substantive impact of AI Legal Reasoning comes to work, and this is characterized by the keyword of "Expressive" denoting that the AI is initially being used as a demonstrative enabler for legal argumentation. Maturing at Level 4, the AI Legal Reasoning is now substantively augmenting legal argumentation, yet does so only within particular legal domains, thus this is characterized as being "Domain Fluency" in its impact. Upon Level 5, encompassing all legal domains, the AI Legal Reasoning has now infused across all legal argumentation and characterized as now being "Full Fluency" in its scope and velocity. Finally, at Level 6, the superhuman AI Legal Reasoning, the advent of micro-directives would be considered "Meta-Fluency," though keep in mind that Level 6 is a speculative notion and it is not clear as to what the superhuman capacity would bring forth.

To reiterate and clarify, these depictions are not prescriptive and do not intend to predict what will happen, and instead are a form of taxonomy to depict and describe what might happen and provide an ontological means to understand such phenomena if it should so arise.

## 4 Additional Considerations and Future Research

As earlier indicated, legal argumentation is a vital cornerstone of justice, underpinning an adversarial form of law, and extensive research has attempted to augment or undertake legal argumentation via the use of computer-based automation including Artificial Intelligence (AI). AI advances in Natural Language Processing (NLP) and Machine Learning (ML) have especially furthered the capabilities of leveraging AI for aiding legal professionals, doing so in ways that are modeled here as *CARE*, namely Crafting, Assessing, Refining, and Engaging in legal argumentation.

In addition to AI-enabled legal argumentation serving to augment human-based lawyering, an aspirational goal of this multi-disciplinary field consists of ultimately achieving autonomously effected human-equivalent legal argumentation. As such, an innovative meta-approach has been proposed to apply the Levels of Autonomy (LoA) of AI Legal Reasoning (AILR) to the maturation of AI and Legal Argumentation (AILA), proffering a new means of gauging progress in this ever-evolving and rigorously sought domain.

Future research is needed to explore in greater detail the manner and means by which AI-enablement will occur, along with the potential for adverse consequences encompassing AI-powered legal argumentation. If such AI Legal Argumentation (AILA) is to be productively adopted, the full gamut of legal, economic, societal, and technological ramifications need to be sufficiently examined.

**About the Author**

Dr. Lance Eliot is the Chief AI Scientist at Techbrium Inc. and a Stanford Fellow at Stanford University in the CodeX: Center for Legal Informatics. He previously was a professor at the University of Southern California (USC) where he headed a multi-disciplinary and pioneering AI research lab. Dr. Eliot is globally recognized for his expertise in AI and is the author of highly ranked AI books and columns.

**References**

1. Alarie, Benjamin, and Anthony Niblett, Albert Yoon (2017). "Regulation by Machine," Volume 24, Journal of Machine Learning Research.

2. Ashley, Kevin, and Karl Branting, Howard Margolis, and Cass Sunstein (2001). "Legal Reasoning and Artificial Intelligence: How Computers 'Think' Like Lawyers," Symposium: Legal Reasoning and Artificial Intelligence, University of Chicago Law School Roundtable.

3. Atkinson, Katie, and Pietro Baroni, Massimiliano Giacomin, Anthony Hunter, Henry Prakken, Chris Reed, Guillermo Simari, Matthias Thimm, Serena Villata (2017). "Toward Artificial Argumentation," AAAI AI Magazine.

4. Baker, Jamie (2018). "A Legal Research Odyssey: Artificial Intelligence as Disrupter," Law Library Journal.

5. Bar-Haim, Roy, and Lilach Eden, Dan Lahav, Roni Friedman, Noam Slonim, Yoav Kantor (2020). "From Arguments to Key Points: Toward Automatic Argument Summarization," Proceedings of the 58th Annual Meeting of the Association for Computational Linguistics.

6. Ben-Ari, Daniel, and D., Frish, Y., Lazovski, A., Eldan, U., & Greenbaum, D. (2016). "Artificial Intelligence in the Practice of Law: An Analysis and Proof of Concept Experiment," Volume 23, Number 2, Richmond Journal of Law & Technology.

7. Bench-Capon, T. (2020). "Before and After Dung: Argumentation in AI and Law," Volume 11, Argument & Computation.

8. Bench-Capon, and Givoanni Sartor (2003). "A Model of Legal Reasoning with Cases Incorporating Theories and Values," November 2013, Artificial Intelligence.

9. Bench-Capon, T., and Paul Dunne (2007). "Argumentation in Artificial Intelligence," Volume 171, Artificial Intelligence.

10. Bilu, Yonatan and Ariel Gera, Daniel Hershcovich, Benjamin Sznajder, Dan Lahav, Guy Moshkowich, Anael Malet, Assaf Gavron, Noam Slonim (2019). "Argument Invention from First Principles," Proceedings of the 57th Annual Meeting of the Association for Computational Linguistics.

11. Braithwaite, John (2002). "Rules and Principles: A Theory of Legal Certainty," Volume 27, Australian Journal of Legal Philosophy.

12. Buchanan, Bruce, and Thomas Headrick (1970). "Some Speculation about Artificial Intelligence and Legal Reasoning," Volume 23, Stanford Law Review.

13. Casey, Anthony, and Anthony Niblett (2016). "Self-Driving Laws," Volume 429, University of Toronto Law Journal.

14. Chagal-Feferkorn, Karni (2019). "Am I An Algorithm or a Product: When Products Liability Should Apply to Algorithmic Decision-Makers," Stanford Law & Policy Review.

15. Chen, Daniel (2019). "Machine Learning and the Rule of Law," in Law as Data: Computation, Text, and The Future of Legal Analysis (Michael A. Livermore and Daniel N. Rockmore eds.).

16. Coglianese, Cary, and David Lehr (2017). "Rulemaking by Robot: Administrative Decision Making in the Machine-Learning Era," Volume 105, Georgetown Law Journal.

17. D'Amato, Anthony (2010). "Legal Uncertainty," Northwestern University School of Law, Faculty Working Papers.

18. Dung, Phan (1993). "On the Acceptability of Arguments and its Fundamental Role in Nonmonotonic Reasoning and Logic Programming," Proceedings of the 13th International Joint Conference on Artificial Intelligence.

19. Eliot, Lance (2016). AI Guardian Angels for Deep AI Trustworthiness. LBE Press Publishing.

20. Eliot, Lance (2020). "The Neglected Dualism of Artificial Moral Agency and Artificial Legal Reasoning in AI for Social Good." Harvard University, Harvard Center for Research on Computation and Society, AI for Social Good Conference, July 21, 2020.

21. Eliot, Lance (2020). AI and Legal Reasoning Essentials. LBE Press Publishing.

22. Eliot, Lance (2019). Artificial Intelligence and LegalTech Essentials. LBE Press Publishing.

23. Eliot, Lance (2020). "FutureLaw 2020 Showcases How Tech is Transforming The Law, Including the Impacts of AI," April 16, 2020, Forbes.

24. Eliot, Lance (2020). "An Ontological AI-and-Law Framework for the Autonomous Levels of AI Legal Reasoning," Cornell University arXiv. https://arxiv.org/abs/2008.07328

25. Eliot, Lance (2020). "Turing Test and the Practice of Law: The Role of Autonomous Levels of AI Legal Reasoning," Cornell University arXiv. https://arxiv.org/abs/2008.07743

26. Eliot, Lance (2020). "Multidimensionality of the Legal Singularity: The Role of Autonomous Levels of AI Legal Reasoning," Cornell University arXiv. https://arxiv.org/abs/2008.10575

27. Eliot, Lance (2020). "Authorized and Unauthorized Practices of Law: The Role of Autonomous Levels of AI Legal Reasoning," Cornell University arXiv. https://arxiv.org/abs/2008.09507

28. Eliot, Lance (2020). "An Impact Model of AI on the Principles of Justice: Encompassing the Autonomous Levels of AI Legal Reasoning," Cornell University arXiv. https://arxiv.org/abs/2008.12615

29. Eliot, Lance (2020). "Robustness and Overcoming Brittleness of AI-Enabled Legal Micro-Directives," Cornell University arXiv. https://arxiv.org/abs/2009.02243

30. Eliot, Lance (2018). "Singularity and AI," July 10, 2018, AI Trends.

31. Feteris E.T. (1999) MacCormick's Theory of the Justification of Legal Decisions. In: Fundamentals of Legal Argumentation. Argumentation Library, vol 1. Springer, Dordrecht.

32. Feteris, Eveline (1999). Fundamentals of Legal Argumentation. Academic Press.

33. Feteris, Eveline, and Harm Kloosterhuis (2009). "The Analysis and Evaluation of Legal Argumentation: Approaches from Legal Theory and Argumentation Theory," Volume 16, Studies in Logic, Grammar and Rhetoric.

34. Gardner, Anne (1987). Artificial Intelligence and Legal Reasoning. MIT Press.

35. Genesereth, Michael (2009). "Computational Law: The Cop in the Backseat," Stanford Center for Legal Informatics, Stanford University.

36. Greenwood, Katie, T. Bench-Capon, and P. McBurney (2003). "Towards a Computational Account of Persuasion in Law," Proceedings of the 9th International Conference on Artificial Intelligence and Law.

37. Habernal, Ivan, and Iryna Gurevych (2017). "Argumentation Mining in User-Generated Web Discourse," Volume 3, Number 1, Computational Linguistics.

38. Hage, Jaap (2000). "Dialectical Models in Artificial Intelligence and Law," Artificial Intelligence and Law.

39. Hermann, Donald (1985). "Legal Reasoning as Argumentation," Volume 12, Northern Kentucky Law Review.

40. Huhn, Wilson (2000). "Teaching Legal Analysis Using a Pluralistic Model of Law," Volume 36, Number 3, Gonzaga Law Review.

41. Huhn, Wilson (2003). "The Stages of Legal Reasoning: Formalism, Analogy, and Realism," Volume 48, Villanova Law Review.

42. Huhn, Wilson (2007). The Five Types of Legal Argument. Carolina Academic Press.

43. Kaplow, Louis (1992). "Rules Versus Standards: An Economic Analysis," Volume 42, Duke Law Journal.

44. Ketcham, Victor (1917). The Theory and Practice of Argumentation and Debate. BiblioLife.

45. Llewellyn, Karl (1950). "Remarks on the Theory of Appellate Decision and the Rules or Canons About How Statutes Are to Be Construed," Volume 3, Number 4, Vanderbilt Law Review.

46. MacCormick, Neil (1978). Legal Reasoning and Legal Theory.

47. MacEwan, Elias (1898). The Essentials of Argumentation. D. C. Heath Publishers.

48. Markou, Christopher, and Simon Deakin (2020). "Is Law Computable? From Rule of Law to Legal Singularity," May 4, 2020, SSRN, University of Cambridge Faculty of Law Research Paper.

49. Marshall, Catherine (1989). "Representing the Structure of a Legal Argument," Proceedings of the 2nd ICAIL.

50. McCarty, Thorne (1977). "Reflections on TAXMAN: An Experiment in Artificial Intelligence and Legal Reasoning," January 1977, Harvard Law Review.

51. McGinnis, John, and Russell G. Pearce (2014). "The Great Disruption: How Machine Intelligence Will Transform the Role of Lawyers in the Delivery of Legal Services," Volume 82, Number 6, Fordham Law Review.

52. McGinnis, John, and Steven Wasick (2015). "Law's Algorithm," Volume 66, Florida Law Review.

53. Mnookin, Robert, and Lewis Kornhauser (1979). "Bargaining in the Shadow of the Law," Volume 88, Number 5, April 1979, The Yale Law Review.

54. Mowbray, Andrew, and Philip Chung, Graham Greenleaf (2019). "Utilising AI in the Legal Assistance Sector," LegalAIIA Workshop, ICAIL, June 17, 2019, Montreal, Canada.

55. Prakken, Henry (1995). "From Logic to Dialectics in Legal Argument," Proceedings of the 5th International Conference on Artificial Intelligence and Law.

56. Prakken, Henry, and Giovanni Sartor (2015). "Law and Logic: A Review from an Argumentation Perspective," Volume 227, Artificial Intelligence.

57. Reed, Chris, and Glenn Rowe (2004). "Araucaria: Software for Argument Analysis, Diagramming, and Representation," Volume 13, Number 4, International Journal on Artificial Intelligence Tools.

58. Reinbold, Patric (2020). "Taking Artificial Intelligence Beyond the Turing Test," Volume 20, Wisconsin Law Review.

59. Remus, Dana, and Frank Levy, "Can Robots be Lawyers? Computers, Robots, and the Practice of Law," Volume 30, Georgetown Journal of Legal Ethics.

60. Rich, Michael (2016). "Machine Learning, Automated Suspicion Algorithms, and the Fourth Amendment," Volume 164, University of Pennsylvania Law Review.

61. Rissland, Edwina (1990). "Artificial Intelligence and Law: Stepping Stones to a Model of Legal Reasoning," Yale Law Journal.

62. SAE (2018). Taxonomy and Definitions for Terms Related to Driving Automation Systems for On-Road Motor Vehicles, J3016-201806, SAE International.

63. Sunstein, Cass (2001). "Of Artificial Intelligence and Legal Reasoning," University of Chicago Law School, Working Papers.

64. Sunstein, Cass, and Kevin Ashley, Karl Branting, Howard Margolis (2001). "Legal Reasoning and Artificial Intelligence: How Computers 'Think' Like Lawyers," Symposium: Legal Reasoning and Artificial Intelligence, University of Chicago Law School Roundtable.

65. Surden, Harry (2014). "Machine Learning and Law," Washington Law Review.

66. Surden, Harry (2019). "Artificial Intelligence and Law: An Overview," Summer 2019, Georgia State University Law Review.

67. Susskind, Richard (2019). Online Courts and the Future of Justice. Oxford University Press.

68. Toulmin, Stephen (1979). An Introduction to Reasoning. MacMillian Press.

69. van Eemeren, Frans and Bart Garssen, Erik Krabbe, Francisca Snoeck Henkemans, Bart Verheij, and Jean Wagemans (2014). Handbook of Argumentation Theory. Springer Berlin.

70. Villata, Serena (2018). "Artificial Argumentation for Humans," Proceedings of the Twenty-Seventh International Joint Conference on Artificial Intelligence.

71. Volokh, Eugne (2019). "Chief Justice Robots," Volume 68, Duke Law Journal.

72. Waltl, Bernhard, and Roland Vogl (2018). "Explainable Artificial Intelligence: The New Frontier in Legal Informatics," February 2018, Jusletter IT 22, Stanford Center for Legal Informatics, Stanford University.

73. Wolfram, Stephen (2018). "Computational Law, Symbolic Discourse, and the AI Constitution," in Data-Driven Law: Data Analytics and the New Legal Services (Edward J. Walters ed.)

**Figure A-1**

## AI & Law: Levels of Autonomy For AI Legal Reasoning (AILR)

| Level | Descriptor | Examples | Automation | Status |
|---|---|---|---|---|
| 0 | No Automation | Manual, paper-based (no automation) | None | De Facto - In Use |
| 1 | Simple Assistance Automation | Word Processing, XLS, online legal docs, etc. | Legal Assist | Widely In Use |
| 2 | Advanced Assistance Automation | Query-style NLP, ML for case prediction, etc. | Legal Assist | Some In Use |
| 3 | Semi-Autonomous Automation | KBS & ML/DL for legal reasoning & analysis, etc. | Legal Assist | Primarily Prototypes & Research Based |
| 4 | AILR Domain Autonomous | Versed only in a specific legal domain | Legal Advisor (law fluent) | None As Yet |
| 5 | AILR Fully Autonomous | Versatile within and across all legal domains | Legal Advisor (law fluent) | None As Yet |
| 6 | AILR Superhuman Autonomous | Exceeds human-based legal reasoning | Supra Legal Advisor | Indeterminate |

*Figure 1: AI & Law - Autonomous Levels by Rows*     *Source Author: Dr. Lance B. Eliot*   V1.3

**Figure A-2**

| | Level 0 | Level 1 | Level 2 | Level 3 | Level 4 | Level 5 | Level 6 |
|---|---|---|---|---|---|---|---|
| **Descriptor** | No Automation | Simple Assistance Automation | Advanced Assistance Automation | Semi-Autonomous Automation | AILR Domain Autonomous | AILR Fully Autonomous | AILR Superhuman Autonomous |
| **Examples** | Manual, paper-based (no automation) | Word Processing, XLS, online legal docs, etc. | Query-style NLP, ML for case prediction, etc. | KBS & ML/DL for legal reasoning & analysis, etc. | Versed only in a specific legal domain | Versatile within and across all legal domains | Exceeds human-based legal reasoning |
| **Automation** | None | Legal Assist | Legal Assist | Legal Assist | Legal Advisor (law fluent) | Legal Advisor (law fluent) | Supra Legal Advisor |
| **Status** | De Facto – In Use | Widely In Use | Some In Use | Primarily Prototypes & Research-based | None As Yet | None As Yet | Indeterminate |

*Figure 2: AI & Law - Autonomous Levels by Columns*     *Source Author: Dr. Lance B. Eliot*     V1.3

**Figure B-1**

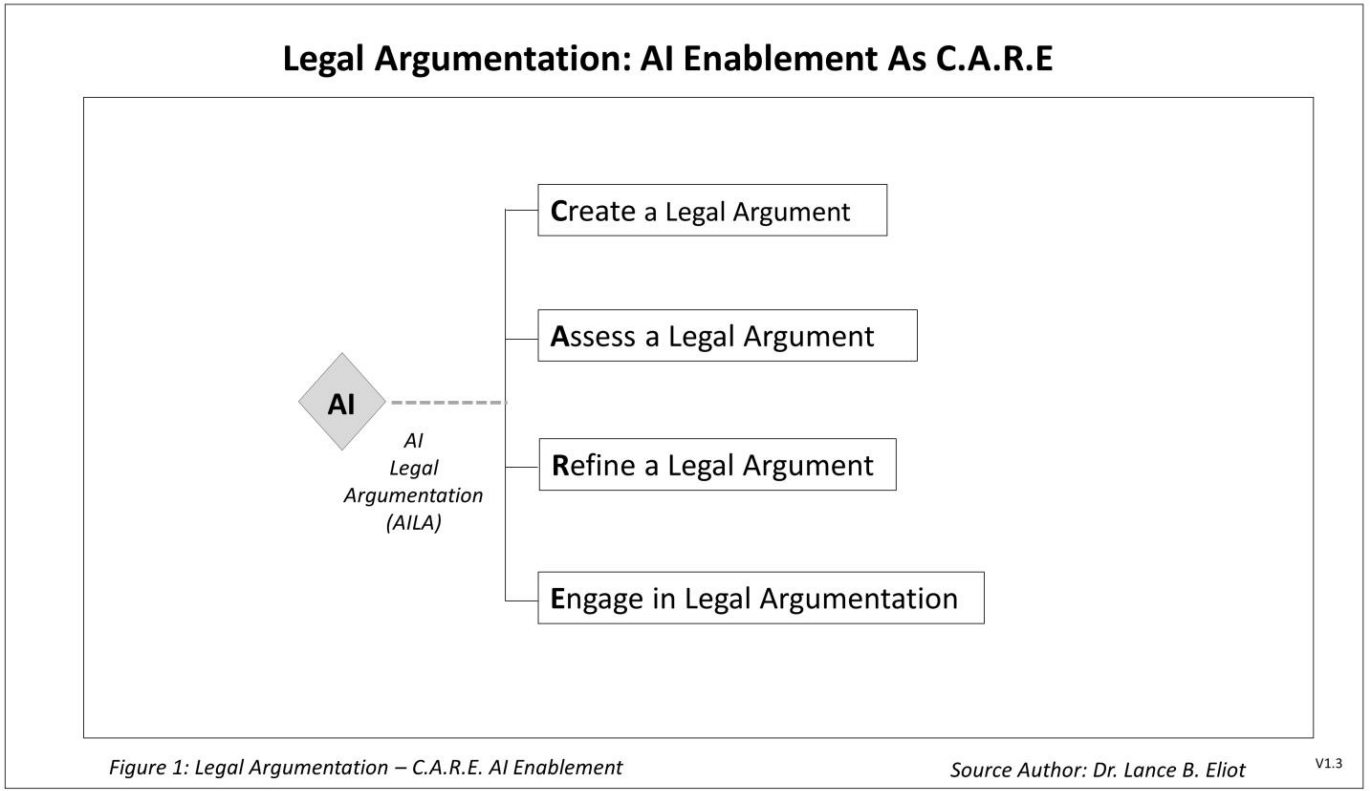

Figure 1: Legal Argumentation – C.A.R.E. AI Enablement  Source Author: Dr. Lance B. Eliot   V1.3

**Figure B-2**

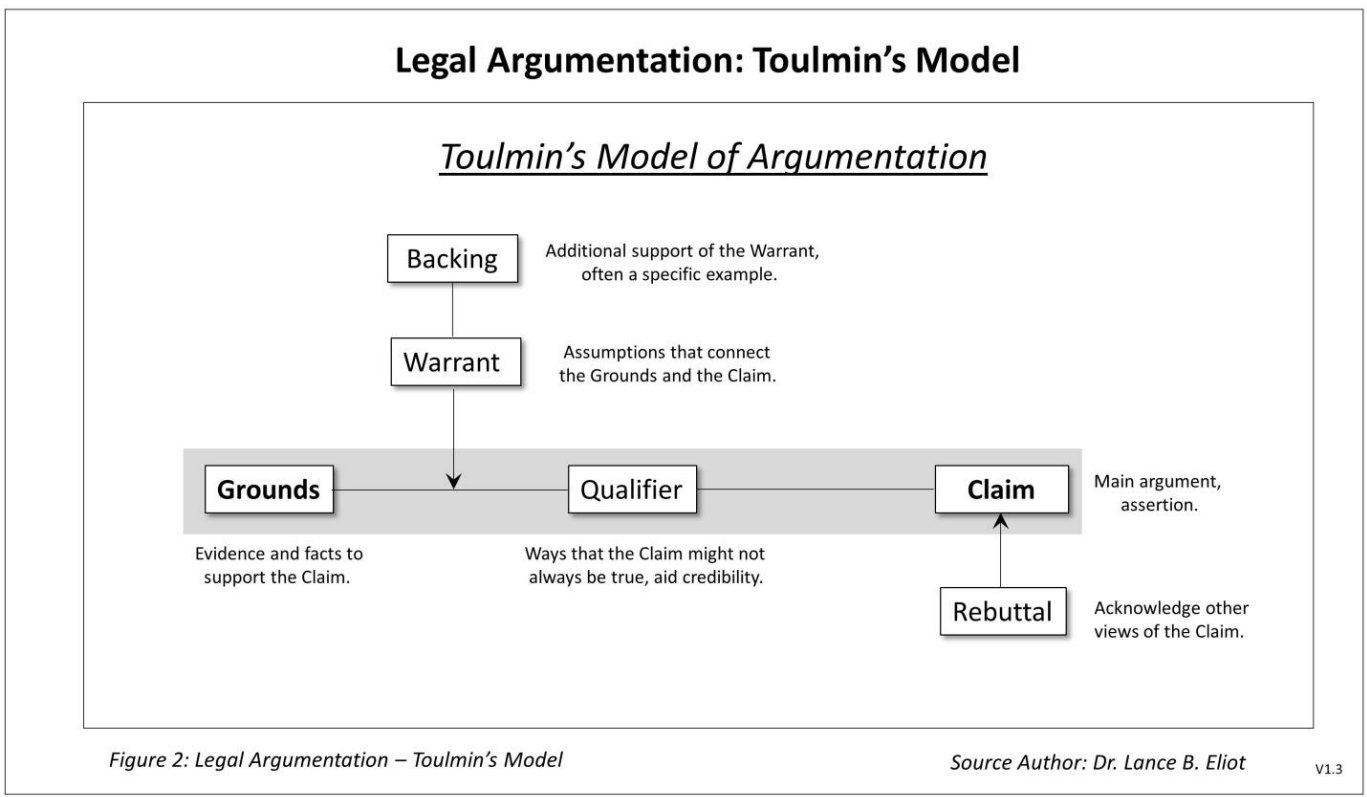

**Figure B-3**

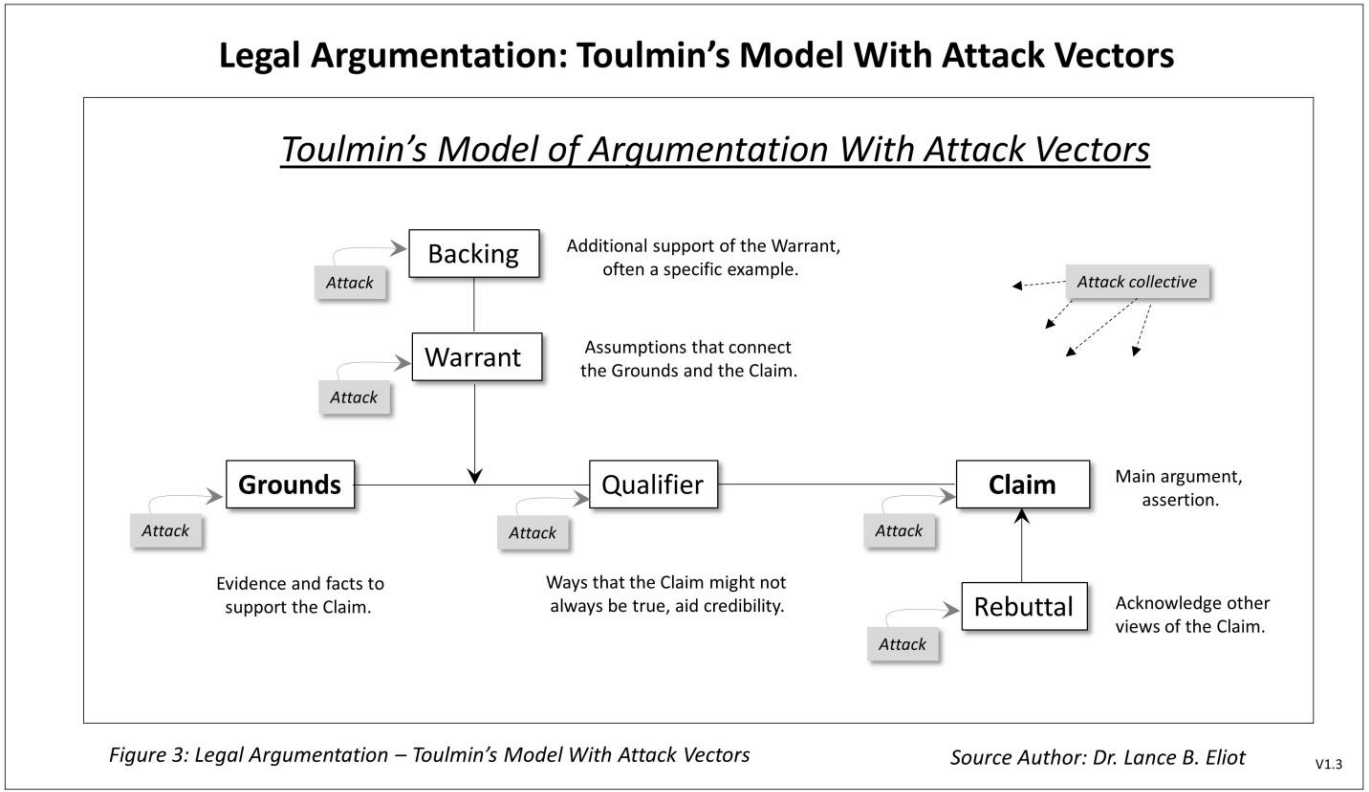

**Figure B-4**

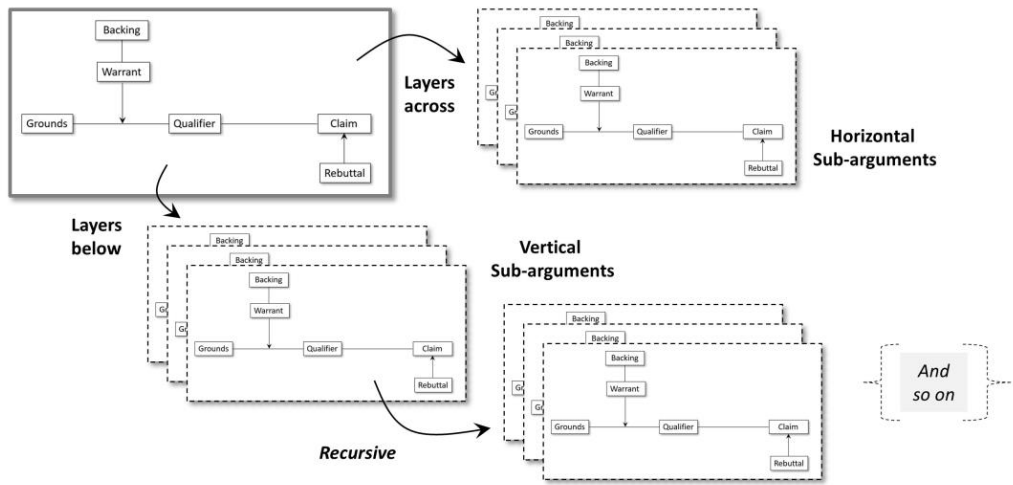

Figure 4: Legal Argumentation – Sub-Arguments at Horizontal and Vertical Layers

**Figure B-5**

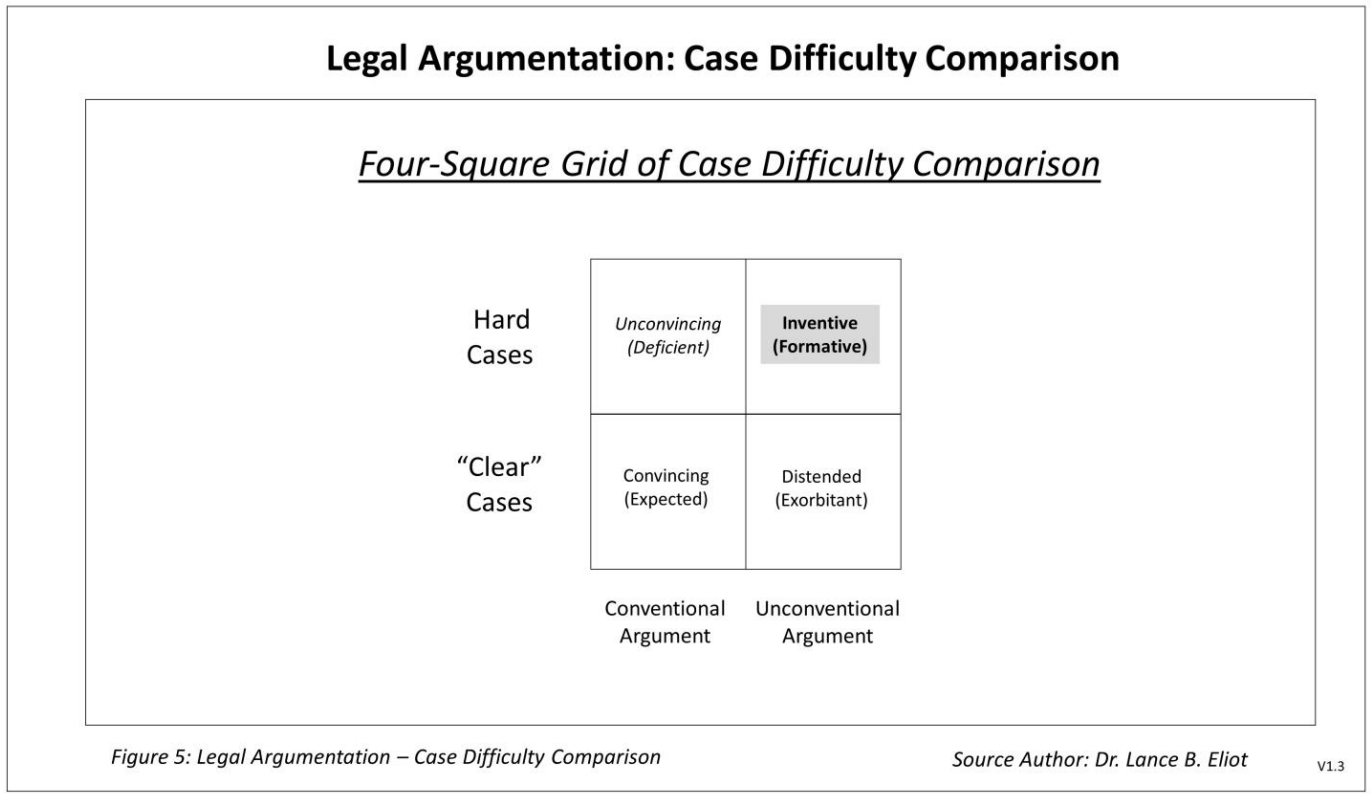

*Figure 5: Legal Argumentation – Case Difficulty Comparison*   *Source Author: Dr. Lance B. Eliot*   V1.3

**Figure B-6**

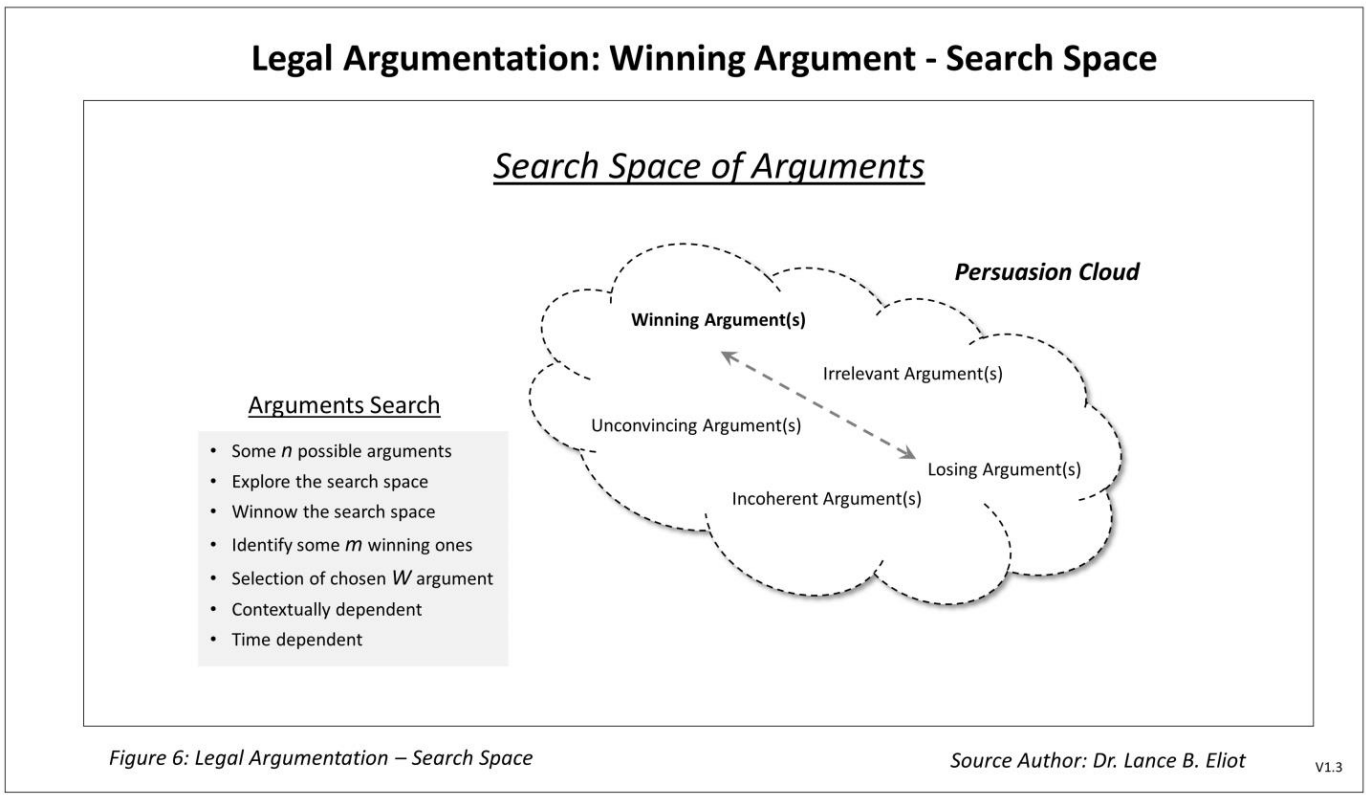

Figure 6: Legal Argumentation – Search Space

**Figure B-7**

## Legal Argumentation: Levels of Autonomy For AI Legal Reasoning (AILR)

|  | Level 0 | Level 1 | Level 2 | Level 3 | Level 4 | Level 5 | Level 6 |
|---|---|---|---|---|---|---|---|
| **Descriptor** | No Automation | Simple Assistance Automation | Advanced Assistance Automation | Semi-Autonomous Automation | AILR Domain Autonomous | AILR Fully Autonomous | AILR Superhuman Autonomous |
| **Examples** | Manual, paper-based (no automation) | Word Processing, XLS, online legal docs, etc. | Query-style NLP, ML for case prediction, etc. | KBS & ML/DL for legal reasoning & analysis, etc. | Versed only in a specific legal domain | Versatile within and across all legal domains | Exceeds human-based legal reasoning |
| **Automation** | None | Legal Assist | Legal Assist | Legal Assist | Legal Advisor (law fluent) | Legal Advisor (law fluent) | Supra Legal Advisor |
| **Status** | De Facto – In Use | Widely In Use | Some In Use | Primarily Prototypes & Research-based | None As Yet | None As Yet | Indeterminate |
| **AI-Enabled Legal Argumentation** | n/a | Mechanistic (Low) | Mechanistic (High) | Expressive | Domain Fluency | Full Fluency | Meta-Fluency |

*Figure 7: AI Legal Argumentation (AILA) - Autonomous Levels of AILR by Columns*  Source Author: Dr. Lance B. Eliot

V1.3